\documentclass[sigconf,natbib=true,anonymous=false]{acmart}
\usepackage{amsmath}
\usepackage{algorithm}
\usepackage{algpseudocode}
\usepackage{enumitem}
\usepackage{tabularx}

\AtBeginDocument{%
  }

\setcopyright{acmlicensed}
\copyrightyear{2025}
\acmYear{2025}
\acmDOI{XXXXXXX.XXXXXXX}
\acmConference[SIGIR '25]{the
48th International ACM SIGIR Conference on Research and Development in Information Retrieval}{July 13--18, 2025}{Padova, Italy}

\begin{document}

\title{Query Attribute Modeling: Improving search relevance with Semantic Search and Meta Data Filtering}

\author{Karthik Menon}
\authornote{Equal Contribution}
\email{karthik@traversaal.ai}
\affiliation{%
  \institution{traversaal.ai}
  \country{USA}
}

\author{Batool Arhamna Haider}
\email{batool@traversaal.ai}
\affiliation{%
  \institution{traversaal.ai}
  \country{USA}
}

\author{Muhammad Arham}
\authornotemark[1]
\email{muhammad.arham@traversaal.ai}
\affiliation{%
  \institution{traversaal.ai}
  \country{USA}
}

\author{Kanwal Mehreen}
\email{kanwalmehreen2000@gmail.com}
\affiliation{%
  \institution{traversaal.ai}
  \country{USA}
}

\author{Ram Mohan Rao Kadiyala}
\email{ram@traversaal.ai}
\affiliation{%
  \institution{traversaal.ai}
  \country{USA}
}

\author{Hamza Farooq}
\email{hamza@traversaal.ai}
\affiliation{%
  \institution{traversaal.ai}
  \country{USA}
}

\renewcommand{\shortauthors}{Menon et al.}

\begin{abstract}
This study introduces Query Attribute Modeling (QAM), a hybrid framework that enhances search precision and relevance by decomposing open text queries into structured metadata tags and semantic elements. QAM addresses traditional search limitations by automatically extracting metadata filters from free-form text queries, reducing noise and enabling focused retrieval of relevant items.

Experimental evaluation using the Amazon Toys Reviews dataset (10,000 unique items with 40,000+ reviews and detailed product attributes) demonstrated QAM's superior performance, achieving a mean average precision at 5 (mAP@5) of 52.99\%. This represents significant improvement over conventional methods, including BM25 keyword search, encoder-based semantic similarity search, cross-encoder re-ranking, and hybrid search combining BM25 and semantic results via Reciprocal Rank Fusion (RRF). The results establish QAM as a robust solution for Enterprise Search applications, particularly in e-commerce systems.
\end{abstract}

\maketitle

\section{Introduction}
The evolution of search engines has progressed from basic retrieval systems to advanced models capable of understanding context and semantics. In its infancy, search engines were primarily concerned with the retrieval of information, employing crawling, indexing, and ranking mechanisms to facilitate access to indexed web pages. Although revolutionary, this initial paradigm lacked the ability to discern contextual relevance and user intent, leading to a search experience that often failed to meet user expectations \cite{10.32628/cseit195115}.

During the mid-1990s \cite{keywordsearch}, a paradigm shift occurred with the emergence of keyword-based search. This approach, epitomized by search engines such as Excite \cite{excite} and WebCrawler \cite{webcrawler}, allowed users to retrieve information based on specific keywords or phrases. However, they exhibit notable weaknesses, such as a lack of understanding of the semantic meaning of queries, which can result in irrelevant results when keywords have multiple meanings \cite{10.1016/j.wpi.2015.02.005}. This shortcoming highlighted the need for more advanced search technologies capable of interpreting the intent and contextual meaning behind queries.

Later era witnessed the emergence of semantic search with methods like Latent Semantic Analysis Theory \cite{landauer1998learning} and TexLexAn \cite{texlexan}. By incorporating natural language processing and machine learning techniques, semantic search systems aimed to provide more accurate and contextually relevant results, marking a departure from simplistic keyword matching paradigms and ushering in a new era of search sophistication and user-centricity. However, semantic search also encounters challenges, including managing language ambiguity, ensuring scalability, and addressing computational overhead, which can result in incomplete or inaccurate results, particularly in complex, real-world scenarios \cite{semanticsearch}.

In recent years, hybrid search\cite{gao2021complement} has emerged as a synergistic fusion of keyword-based precision and semantic contextual understanding. This hybrid approach combines the strengths of both the keyword-based and semantic search approaches, thus enhancing the overall search experience for users. Despite its promise, challenges persist in integrating keyword and semantic results, particularly in scenarios involving complex queries and rich metadata like, \textit{"I am looking for educational toys specifically from LEGO, designed to promote creativity, suitable for children aged 5-8"} and \textit{"Locate a top-rated board game from Hasbro for kids aged 9-12 within a budget of \$40"}.

Against this backdrop of evolving search methodologies, Query Attribute Modeling (QAM) emerges as a new paradigm designed to redefine enterprise search. QAM introduces a novel framework that harmonizes the semantic and keyword-based capabilities, addressing the inherent limitations of existing search systems. By systematically dissecting user queries into structured metadata tags and semantic components, QAM enables a more precise and contextually relevant interpretation of user queries.

The primary objective of this research is to demonstrate how Query Attribute Modeling enhances search precision and relevance based on user open text search. Through detailed experimentation and analysis, we aim to showcase its potential to transform enterprise search by addressing the challenges of scalability, efficiency, and adaptability in handling complex real-world queries. The following sections outline the methodology (Section~\ref{sec:methodology}), experimentation (Section~\ref{sec:experimentation}), and results (Section~\ref{sec:results}), highlighting QAM's effectiveness in meeting the growing demands of modern search technologies.

\section{Methodology}
\label{sec:methodology}
\begin{figure*}[h]
    \centering
    \includegraphics[width=0.85\textwidth]{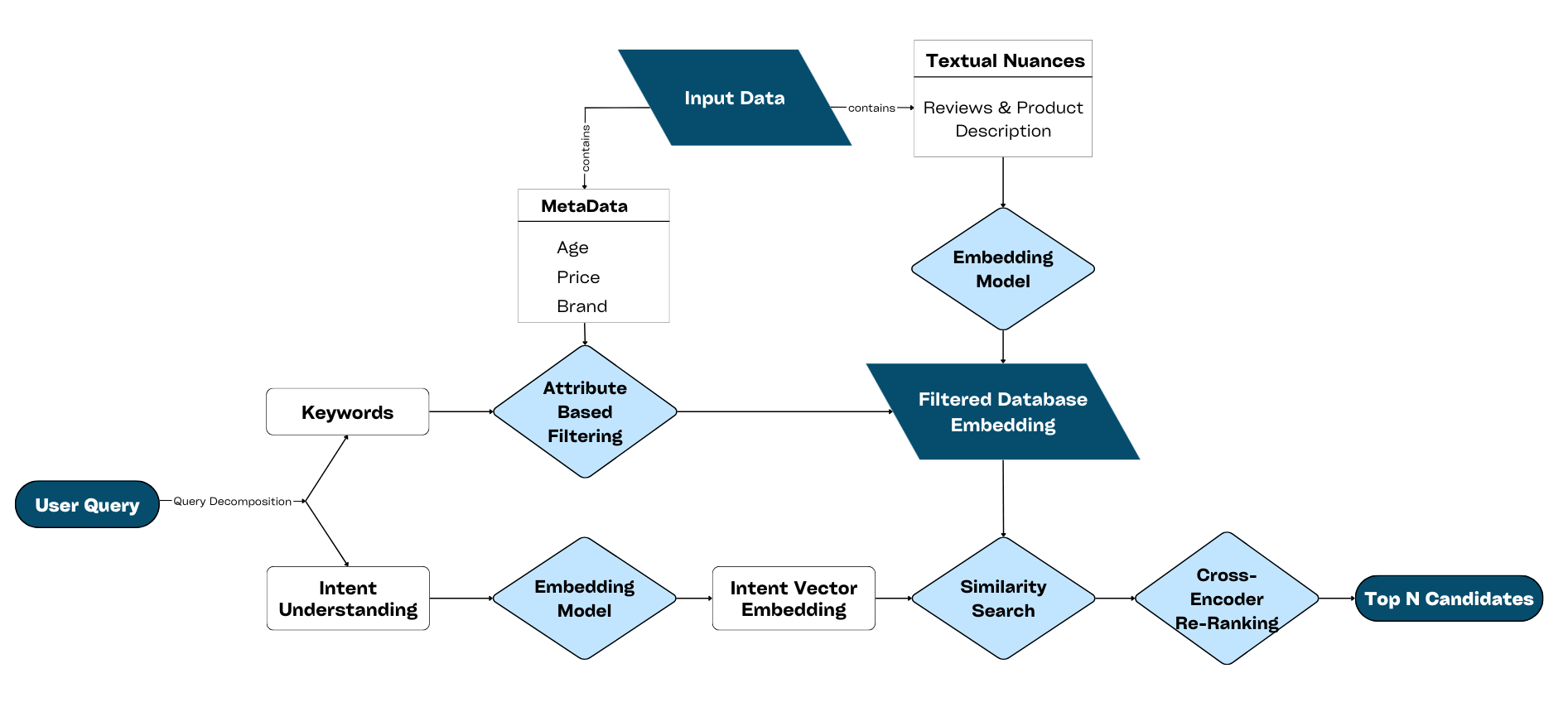}
    \caption{Query Attribute Modeling (QAM) methodology, illustrating the four-stage process of query understanding, metadata filtering, semantic search, and re-ranking of results from the filtered dataset.}
    \label{fig:methodology-diagram}
\end{figure*}

The methodology employed in our research follows a systematic approach to enhance the precision and relevance of search results within the context of Query Attribute Modeling (QAM). It comprises of four distinct steps, each designed to address specific aspects of search refinement and optimization, as shown in Figure \ref{fig:methodology-diagram}.

\begin{algorithm}
\caption{QAM Algorithm}\label{alg:qam}
\begin{algorithmic}
[1]
\Require Query $Q$, Dataset $D$
\State \textbf{Input:} $Q = \texttt{"A long black dress from Zara under \$100"}$
\State \textbf{Output:} Ranked search results $R$
\Statex
\vspace{10pt}
\textbf{Step 1: Query Decomposition}
\State $Q_{\text{metadata}} \gets \text{Extract metadata tags (e.g., color, brand)}$
\State $Q_{\text{semantics}} \gets \text{Extract semantic elements}$
\vspace{8pt}
\Statex
\textbf{Step 2: Metadata Filtering}
\State $D_{\text{filtered}} \gets \{ p \in D \mid p.\text{metadata matches } Q_{\text{metadata}} \}$
\Statex
\vspace{8pt}
\textbf{Step 3: Review Similarity}
\For{\textbf{each product} $p \in D_{\text{filtered}}$}
    \State $p.\text{score} \gets \text{CosSim(Enc}(Q_{\text{semantics}}, p)\text{)}$
\EndFor
\Statex
\vspace{8pt}
\textbf{Step 4: Final Ranking}
\For{\textbf{each product} $p \in D_{\text{filtered}}$}
    \State $p.\text{final\_score} \gets \text{CrossEncoder}(Q, p)$
\EndFor
\State $R \gets \text{Sort}(D_{\text{filtered}}, \text{by}=p.\text{final\_score})$
\Statex
\vspace{5pt}
\Return Top-$N$ results from $R$
\end{algorithmic}
\end{algorithm}

\subsection{Query Decomposition}

The first step focuses on dissecting user queries into two primary components: metadata tags and semantic elements. This decomposition enables the search system to separate explicit user requirements (e.g., ``color'' or ``brand'')) from the deeper contextual meaning of the query. To achieve this, we employ a language model (e.g., GPT-4) \cite{openai2024gpt4ocard}, which excels in parsing complex queries and extracting structured information. 
\begin{itemize}
  \item \textbf{Metadata Tags:} These include structured attributes such as product brand, material, price constraints, and preferred user demographics (e.g., age groups). These tags provide a structured way for filtering datasets effectively.
  \vspace{5pt}
  \item \textbf{Semantic Elements:} These capture the contextual intent of the query, allowing the system to understand implicit preferences and refine results accordingly.
\end{itemize}

\subsection{Metadata Filtering for Enhanced Search Precision}

Building upon the extracted metadata tags, the subsequent step focuses on enhancing search precision by using these tags to filter the dataset and retain only the most relevant items. Metadata attributes such as material, brand, and color play a crucial role in this filtering process. For instance, in a query like "a little black dress," the system utilizes the extracted metadata tag "black" \& "Zara" to exclude irrelevant results, such as dresses of other colors or brands. Similarly, filtering by material and brand ensures that user preferences are prioritized early in the pipeline, reducing computational overhead for subsequent steps. This method enhances both efficiency and precision by eliminating noise from the dataset. Metadata filtering has been shown to be a lightweight yet impactful technique for aligning search results with user intent \cite{10.2196/25440}.

\subsection{Query and Product Description Similarity Search}

This step employs semantic embeddings and cosine similarity to connect user queries with relevant qualitative information in product reviews. Semantic embeddings, generated using advanced models like nomic-embed-text-v1 \cite{nussbaum2024nomicembedtrainingreproducible}, encode the contextual meaning of the query and reviews into vector representations. Cosine similarity is then calculated to measure how well a product aligns with the user's intent. For example, if a query specifies \textit{"suitable for formal events,"} this step prioritizes products with reviews mentioning \textit{"formal occasions"}. By linking the subjective components of the query with qualitative descriptions in the reviews, this step deepens the system’s understanding of user requirements and enhances result relevance. This builds on existing methodologies using contextual review analysis to improve search outcomes \cite{10.1109/itng.2009.133}.

\subsection{Final Ranking}

The final step integrates the outputs of the previous phases to deliver the most relevant results. A cross-encoder model, such as msmarco-MiniLM-L12-en-de-v1 \cite{huggingface2025msmarco}, is employed to compute the final relevance score for each product. Unlike bi-encoders, which generate separate embeddings for queries and products and compute relevance scores based on their similarity, cross-encoders process the query and product together, directly modeling their interaction. This approach allows cross-encoders to capture finer-grained relationships between the query and product, leading to more accurate rankings \cite{10.48550/arxiv.2302.04112}. For each product in the filtered dataset, the cross-encoder computes a final score based on the semantic similarity between the query and product attributes. The results are then sorted by these scores to produce a ranked list of items, ensuring that the most relevant results are prioritized. This step ensures the delivery of highly personalized and contextually relevant search results.

\section{Experimentation}
\label{sec:experimentation}

\subsection{Data}
The experimentation phase utilized the \textit{Amazon Toys Reviews} dataset, which consists of 10,000 unique items with product descriptions and 40,000+ reviews including 15 raw and engineered features. This data set was chosen for its extensive coverage of product reviews, which facilitates a detailed analysis at the review level for each product. 

In addition to reviews, a significant focus was placed on feature extraction from product descriptions. This involved
extracting essential attributes such as brand and required minimum age. To achieve this, advanced text preprocessing techniques were applied, using natural language processing (NLP) libraries such as NLTK and spaCy. These techniques enabled the extraction of pertinent information from the textual descriptions, enriching the dataset with valuable metadata.

To evaluate QAM and its competing methods, a diverse set of 1,000 queries was generated using GPT-4o. These queries were designed to simulate realistic user searches, capturing both explicit requirements (e.g., brand, price, age) and subjective intent (e.g., suitability for specific occasions). Out of the generated queries, 200 high-quality queries were selected for the evaluation dataset to ensure alignment of brand names and attributes with the entries in the original Amazon dataset. Examples include: \textit{"Can I find Playteachers toys for kids aged 6 to 15?"} and \textit{"Looking for a Kaleidoscope toy for my 3-year-old, priced around \$12."} This carefully curated query set provided a robust basis for evaluating QAM’s hybrid approach to address both explicit preferences and contextual query intent.

\vspace{-5pt}
\subsection{Evaluation Setup}
The evaluation involved running each query against five search methods: BM25 keyword-based search\cite{robertson2009probabilistic}, semantic search\cite{semanticsearch}, cross-encoder re-ranking, hybrid search, and QAM. Each method returned the top 10 results, which were annotated for relevance using an LLM (GPT-4).

\textbf{Annotation Process}: The LLM was given both the query and the returned results and was tasked with determining whether each result was relevant. The relevance was based on the following:
\begin{itemize}[noitemsep, topsep=2pt, partopsep=2pt]
  \item Exact match for metadata (e.g., price, brand). For quantitative values including rating and price, we allowed for a 20\% percent complacency between the returned value and the required value to allow flexibility in responses.
  \item Semantic alignment for contextual preferences.
\end{itemize}

\textbf{Scoring Metrics}: The annotated results were evaluated using precision@k (P@k) and mean average precision@k (mAP@k). These metrics captured the accuracy and ranking quality of each method \cite{prec_k_manning}.  

Precision at k (P@k) measures the ratio of relevant items among the top K results, as shown in Equation~\eqref{eq:precision_k}.  

\small
\begin{equation}
    Precision@k = \frac{\text{Relevant Results @k}}{k}
    \label{eq:precision_k}
\end{equation}
\normalsize

Average Precision@K (AP@K) calculates the precision at each rank where a relevant item appears, averaged over all relevant items, as defined in Equation~\eqref{eq:ap_k}.  

\small
\begin{equation}
    AP@K = \frac{1}{\min(K, \text{Total Relevant Items})} \sum_{i=1}^{K} P(i) \cdot rel(i)
    \label{eq:ap_k}
\end{equation}
\normalsize

\begin{figure*}[h]
    \centering
    \includegraphics[width=0.85\textwidth]{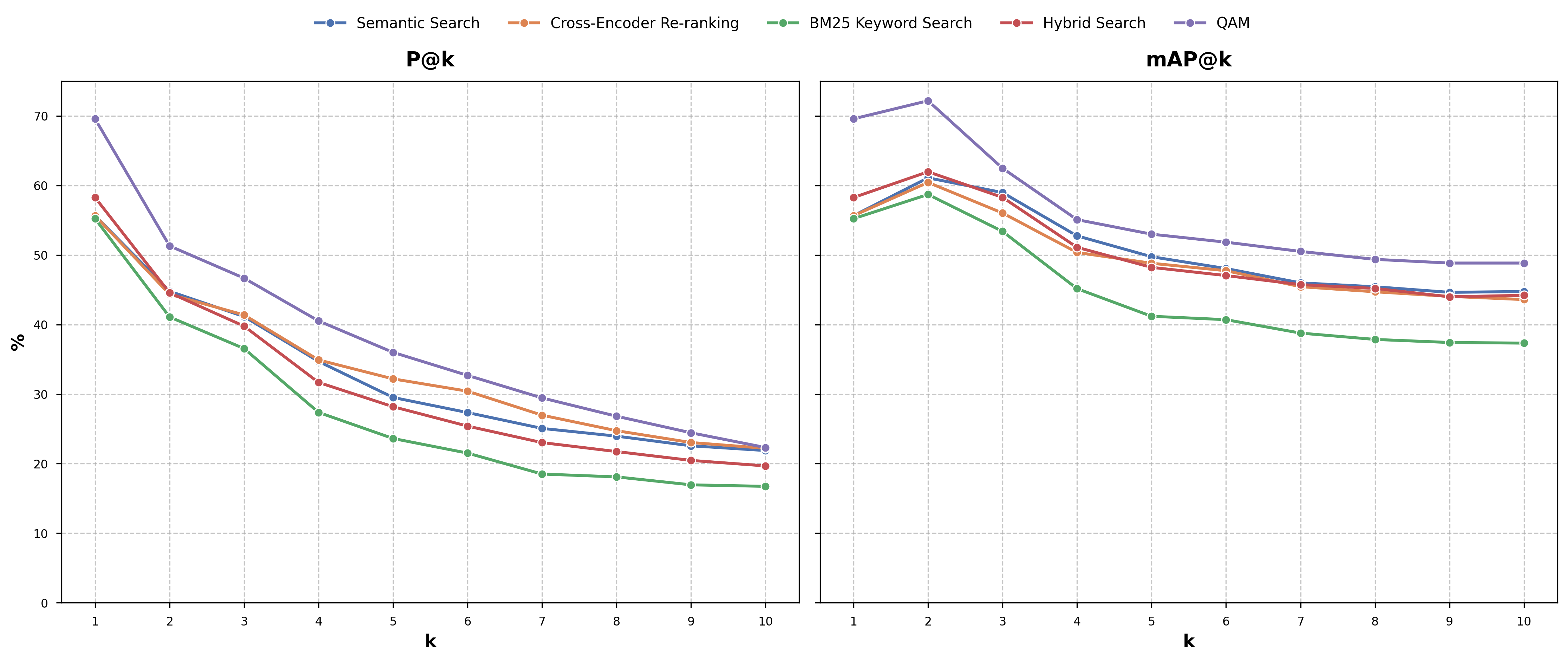}
    \caption{Comparison of Precision and mAP scores for five retrieval methods across varying k values. The y-axis represents the percentage of relevant documents among the top-k results, with QAM outperforming all other methods across all k values.}
    \label{fig:comparison-plot}
\end{figure*}

AP@k score values the ranking of retrieved results, returning a higher score if relevant data points are ranked higher than non-relevant results. Mean Average Precision (mAP@K) computes the mean of AP@K scores across all data samples, providing an aggregate score for all queries, as given in Equation~\eqref{eq:map_k}.  

\small
\begin{equation}
    mAP@K = \frac{1}{N} \sum_{q=1}^{N} AP@K_q
    \label{eq:map_k}
\end{equation}
\normalsize
The use of an LLM as a judge automated the annotation process, reducing human bias and ensuring consistent evaluation standards \cite{10.1002/ev.20556}. Additionally, for complex queries and certain metadata combinations, the QAM search method may significantly reduce the candidate set, yielding fewer than k retrieved results. To ensure an unbiased evaluation, we restricted our analysis to search instances where at least k relevant documents were available. In cases where fewer than k relevant results existed, the missing results were treated as non-relevant, thereby penalizing our approach for failing to retrieve the desired number of relevant documents.

\vspace{-10pt}
\section{Results and Analysis}
\label{sec:results}
The results demonstrate that QAM significantly outperforms traditional search methods. 

In terms of Mean Average Precision at 5 (mAP@5), QAM achieved a score of 52.99\%, which is consistently higher than the scores of the other methods. Specifically, QAM showed a 28.67\% improvement over BM25 keyword-based search (41.19\%), 6.5\% over semantic search (49.75\%), and 8.58\% over cross-encoder reranking (48.81\%). QAM achieved a 9.96\% improvement compared to hybrid search, which combined encoder embeddings and BM25 search results using Reciprocal Rank Fusion (RRF) and scored 48.22\%. Table \ref{tab:map_scores} summarizes the mAP@K scores for all methods.

Furthermore, the comparison of Precision@K, summarized in Table \ref{tab:precision_scores}, across all methods demonstrates the consistent superiority of QAM. Across all values of k (1 to 10), QAM consistently retrieves a higher percentage of relevant results compared to other methods. Figure \ref{fig:comparison-plot} summarizes these findings, illustrating how P@K and mAP@K vary with k. The results indicate that QAM outperforms all other approaches in the Amazon toy data set retrieval task by effectively filtering out irrelevant results prior to searching, thus improving the overall relevance of the retrieved documents.

\begin{table}[h!]
\centering
{\fontsize{8}{10}\selectfont}
\captionsetup{font=footnotesize} 
\captionsetup{skip=3pt} 
\caption{Precision@K Scores Across Methods}
\label{tab:precision_scores}
\renewcommand{\arraystretch}{1.1} 
\setlength{\tabcolsep}{3pt} 
\begin{tabularx}{\columnwidth}{|X|X|X|X|}
\hline
\textbf{Method}             & \textbf{P@3} & \textbf{P@5} & \textbf{P@10} \\ 
\hline
Keyword Search           & 36.55\%      & 23.62\%      & 16.74\%       \\ \hline
Semantic Search        & 41.15\%      & 29.52\%      & 21.89\%       \\ \hline
Re-Ranking        & 41.38\%      & 32.19\%      & 22.21\%       \\ \hline
Hybrid Search        & 39.77\%      & 28.19\%      & 19.68\%       \\ \hline
\textbf{QAM}     & \textbf{46.67\%} & \textbf{36.00\%} & \textbf{22.32\%} \\ \hline
\end{tabularx}
\end{table}

\begin{table}[h!]
\centering
{\fontsize{8}{10}\selectfont}
\captionsetup{font=footnotesize} 
\captionsetup{skip=3pt} 
\caption{Mean Average Precision (mAP@K) Scores}
\label{tab:map_scores}
\renewcommand{\arraystretch}{1.1} 
\setlength{\tabcolsep}{2pt} 
\begin{tabularx}{\columnwidth}{|X|X|X|X|}
\hline
\textbf{Method}  &  \textbf{mAP@3} & \textbf{mAP@5} & \textbf{mAP@10} \\ \hline
Keyword Search                             & 53.39\%        & 41.19\%        & 37.33\%         \\ \hline
Semantic Search                         & 58.97\%        & 49.75\%        & 44.75\%         \\ \hline
Re-Ranking                       & 56.03\%        & 48.81\%        & 43.59\%         \\ \hline
Hybrid Search                       & 58.28\%        & 48.22\%        & 44.2\%         \\ \hline
\textbf{QAM}                   & \textbf{62.47\%} & \textbf{52.99\%} & \textbf{48.84\%} \\ \hline
\end{tabularx}
\end{table}

Thus, QAM outperforms hybrid search and other retrieval methods in scenarios requiring both specificity and contextual understanding. Unlike hybrid search, which combines multiple ranking signals, QAM’s structured approach enhances relevance and retrieval accuracy, making it a more effective solution for modern search challenges.
\section{Conclusion \& Next Steps}
In conclusion, this research introduces Query Attribute Modeling (QAM), an innovative framework for enhancing precision and relevance in search systems. By systematically integrating query decomposition, metadata filtering, and contextual analysis, QAM consistently outperforms traditional keyword-based and semantic search methods.
For the next phase, we aim to enable the Language Model (LLM) API to autonomously identify relevant keyword tags from user queries, eliminating the need for explicit guidance and enhancing the dynamism of our query deconstruction process. Additionally, the integration of powerful vector databases like Qdrant \cite{qdrant_tech} will streamline information retrieval, contributing to a more sophisticated search experience. We intend to address scalability limitations inherent in manual data labeling by scaling our model to standard databases and a wider array of queries, ensuring stability and robustness across diverse datasets. 
\newpage
\bibliographystyle{ACM-Reference-Format}
\bibliography{ref}
\end{document}